\documentclass[aps,pre]{revtex4}
\usepackage{graphicx}
\usepackage{amssymb,amsfonts,amsmath}

\pdfoutput=1

\begin{document}

\title{Characterizing Structure Through Shape Matching and Applications to Self Assembly}

\author{Aaron S. Keys$^1$}
\author{Christopher R. Iacovella$^1$}
\author{Sharon C. Glotzer$^{1,2}$}
\affiliation{$^1$Department of Chemical Engineering and $^2$Department of Materials Science and Engineering \\University of Michigan, Ann Arbor, Michigan 48109-2136}

\date{\today}

\begin{abstract}

Structural quantities such as order parameters and correlation functions are often employed to gain insight into the physical behavior and properties of condensed matter systems.  While standard quantities for characterizing structure exist, often they are insufficient for treating problems in the emerging field of nano and microscale self-assembly, where the structures encountered may be complex and unusual. The computer science field of ``shape matching'' offers a robust solution to this problem by defining diverse methods for quantifying the similarity between arbitrarily complex shapes.  Most order parameters and correlation functions used in condensed matter apply a specific measure of structural similarity within the context of a broader scheme. By substituting shape matching quantities for traditional quantities, we retain the essence of the broader scheme, but extend its applicability to more complex structures.  Here we review some standard shape matching techniques and discuss how they might be used to create highly flexible structural metrics for diverse systems such as self-assembled matter.  We provide three proof-of-concept example problems applying shape matching methods to identifying local and global structures, and tracking structural transitions in complex assembled systems.  The shape matching methods reviewed here are applicable to a wide range of condensed matter systems, both simulated and experimental, provided particle positions are known or can be accurately imaged.

\end{abstract}

\maketitle

\section{Introduction}
The preponderance of new nanometer- and micron-sized colloidal particles of nearly arbitrary shape, composition and interaction has made possible the self-assembly of exquisitely complex structures with potential uses in a variety of technologies \cite{kumacheva,desimone2006,glotzer07, glotzer2005}.  Because material properties and behavior are determined by both the global and local shapes, or patterns, within the self-assembled structure\cite{kumacheva, daniel2004,collier, storhoff, kempa, reinhard}, methods and tools are needed to characterize the salient structural features of the assemblies.  The field of condensed matter physics has traditionally led the way in developing algorithms for characterizing crystal structures and constructing theories to connect these structures to thermodynamics and to overall system properties\cite{flory, onsager, bernal}.  These approaches typically involve constructing structural order parameters and/or correlation functions that can discriminate between different building block arrangements and are well developed for systems of point-like, rod-like and spherical particles\cite{larson, mermin1968, tenwolde96, cna, kamienchiral}.  Examples include nematic and smectic order parameters for systems of rods cite{nematic, smectic, liquidcrystals} and bond order parameters\cite{mermin1968, halperin, snr83, tenwolde96} for 2d and 3d systems of spheres.

However, these functions fail, in many cases, to fully describe the structural complexity of assemblies of more unusual nanocolloids, including those formed from spherical particles\cite{storhoff, akcora},  rod-like particles\cite{nie2007, kempa}, polyhedral particles\cite{amir09, tang2006, zhang2007, zhang2003, twistedribbons, magcubes, dipolecube, bettina}, colloidal molecules\cite{pine2003, pine2005, kraft, siva, burns2008, glotzer07}, patchy spheres\cite{mohwald2005, zhang2004, zhenlidiamond, devries2007, jackson2004, pawar}, arbitrarily-shaped objects\cite{kumacheva, glotzer07}, polymer-tethered nanoparticles \cite{glotzer2005, zhang2003, park, reister, arthi2008, waddon2002, rotello}, and terminal assemblies resembling biological structures\cite{virus, rapaport2004self}.  For example, it is easy to envision that order parameters defined for spherical or rod shaped particles may fail when applied to more complex shaped particles, such as ``Y''  particles or triangular plates\cite{glotzer07}.  As a result of the increased complexity of nano building blocks, there are few ``model problems'' in nano and microscale self-assembly for which generally applicable order parameters can be defined.   The dearth of structural metrics has lead many recent  experimental and computational studies of assembled systems to rely heavily on visual inspection or \textit{ad hoc} analysis for characterizing structures, rather than well established schemes.  This approach is not optimal, since visual inspection can be time consuming and typically less accurate than mathematical analysis, and \textit{ad hoc} analysis can be idiosyncratic, making it difficult to compare structures across independent studies.  The impetus for new structural metrics is also driven by advances in microscopy techniques that allow for the direct imaging of nano and microscale systems, which have greatly extended the range of systems for which detailed structural analysis can potentially be performed.  For example, the tracking of micron-sized colloidal particles in 2d and 3d is now routine\cite{crocker1996, weeks2000, laura,tesfuv2,varadan2003}, and high-fidelity imaging of nanoparticles\cite{coura} and their assemblies\cite{tang2006, talapin, murray} is steadily improving.  Combined with proper image processing techniques, one can extract much information about structure, such as the particle positions\cite{varadan2003, crocker1996} and other key features, providing detailed structural information on par with simulations.  Assuming one can construct order parameters sensitive to these unique building blocks and their assemblies, similar routines can be applied to both experimental and simulated systems, allowing for direct comparison\cite{siva, tang2006}.

\begin{figure*}[h!]
\begin{center}
\includegraphics[width=0.8\textwidth]{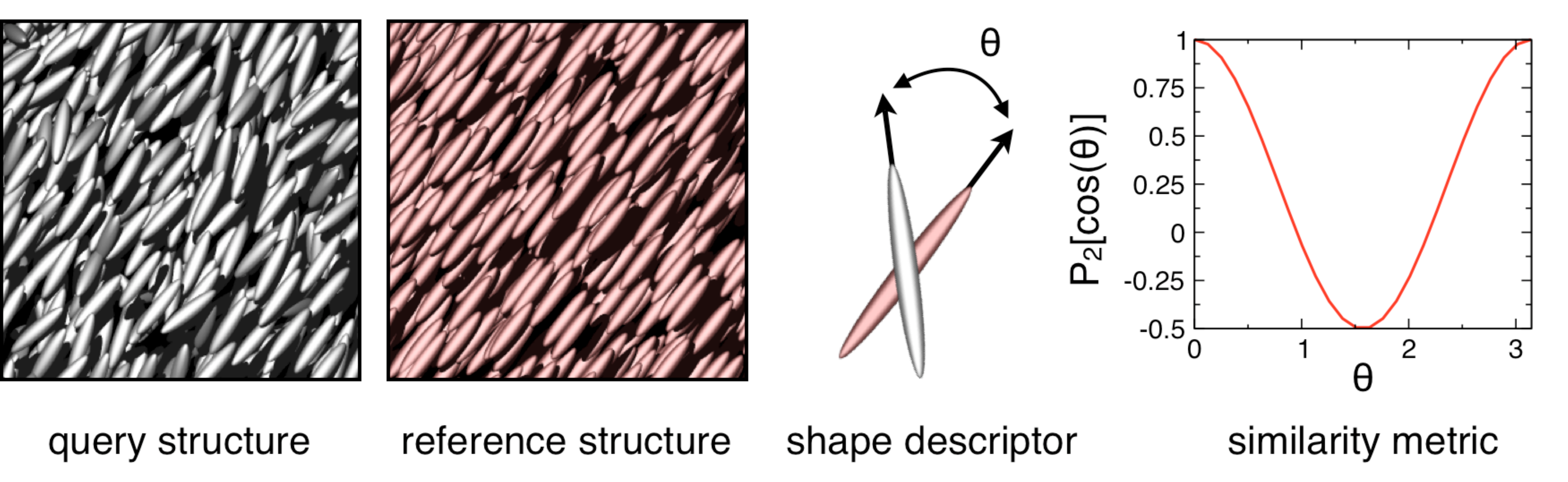}
\caption{Example of an implicit shape matching scheme within the context of a standard order parameter.  The panel depicts the process of computing the nematic order parameter $\bar{P}_2$ for a system of rod-like colloidal ellipsoids that assemble into an aligned ordered phase\cite{mohraz}.  In the language of a shape matching scheme (see section~\ref{sec:shapematching}), the colloidal system acts as a ``query structure'' that we wish to characterize.  An ideal system for which the rods are all oriented along the average global director acts as an implicit ``reference structure.''  The local values of the angles $\theta$ between rods in the query structure and reference structure act as ``shape descriptors.''  The Legendre polynomial $P_2$ acts as a ``similarity metric.''  The global nematic order parameter $\bar{P}_2$ is computed by averaging over local values of $P_2[\cos(\theta)]$.}
\label{fig:nematic}
\end{center}
\end{figure*}

Analysis techniques from the computer science field of ``shape matching'' offer a potentially powerful solution to the problem of creating general structural metrics for these systems.  Shape matching involves defining general structural metrics that can be used to measure the degree of similarity between diverse shapes.  Such similarity measures can be applied within the context of traditional condensed matter order parameter and correlation function schemes to obtain analogous quantities for more complex structures.  This is possible because, in practice, most standard structural characterization schemes include an implicit concept of matching or shape similarity; that is, the schemes typically measure the degree to which a structure of interest matches another (often ideal) structure.  As a familiar example, consider the standard nematic order parameter which gives an optimal value of $1$ when the rod-like particles within the system are perfectly aligned, and $0$ if the rods have random orientations.  In this case, the order parameter measures the degree to which the local arrangement of rods in the system, described mathematically by the angles between neighboring rods, matches with an ideal reference system with perfect alignment (see Fig.~\ref{fig:nematic}).  Other structural characterization schemes and spatial or temporal correlation functions involve similar underlying concepts of matching.  As we will discuss, by modifying these schemes to use shape matching methods, we retain their overall physical insight, but gain the ability to apply them to complex structures.  Although we focus exclusively on simulated assembled systems here, these types of methods are general enough that they can be applied to particle systems in general, provided that the particle positions and or orientations can be determined or imaged.  Examples of systems, both experimental and simulated, to which shape matching methods can potentially be applied include but are not limited to nanoparticle superlattices created from mixtures of spherical and/or non-spherical nanoparticles\cite{murray,talapin}, microphase separated systems, such as tethered nanoparticles and block copolymers that form crystalline and quasicrystalline domains\cite{ditethered, dotera}, colloidal ionic crystals\cite{vanblaaderan}, dense colloids\cite{weeks2000} and granular matter\cite{liu98, abate}. 

This review is organized as follows.  In section~\ref{sec:shapematching}, we review shape matching methods from the literature, restricting our scope to methods that we believe are most immediately applicable to assembled systems.  We describe how representative shapes can be extracted from particle systems, review the shape descriptors that are best suited to describe these shapes numerically, and show how they can be compared quantitatively.  In section~\ref{sec:applications}, we apply a prototype shape matching scheme to three representative example problems from simulations of self-assembly.   Our examples include identifying global structures in a microphase-separating system of polymer-tethered nanospheres\cite{ditethered}, detecting local icosahedral clusters in a fluid of hard tetrahedral particles\cite{amir09}, and tracking the twisting of a helical sheet formed from polymer-tethered nanorods\cite{trunghelix}.  In section~\ref{sec:future}, we suggest new applications for shape matching methods, including constructing correlation functions, measuring local crystal grains and crystal defects, devising guided computer algorithms to map parameter spaces and search for target structures, and grouping and classifying structures based on particular structural features.  To aid in the development and dissemination of new structural analysis methods based on shape matching techniques, we provide accompanying software and examples via the web\cite{smwebsite}.

\section{Shape Matching}
\label{sec:shapematching}

\begin{figure*}[h!]
\begin{center}
\includegraphics[width=0.6\textwidth]{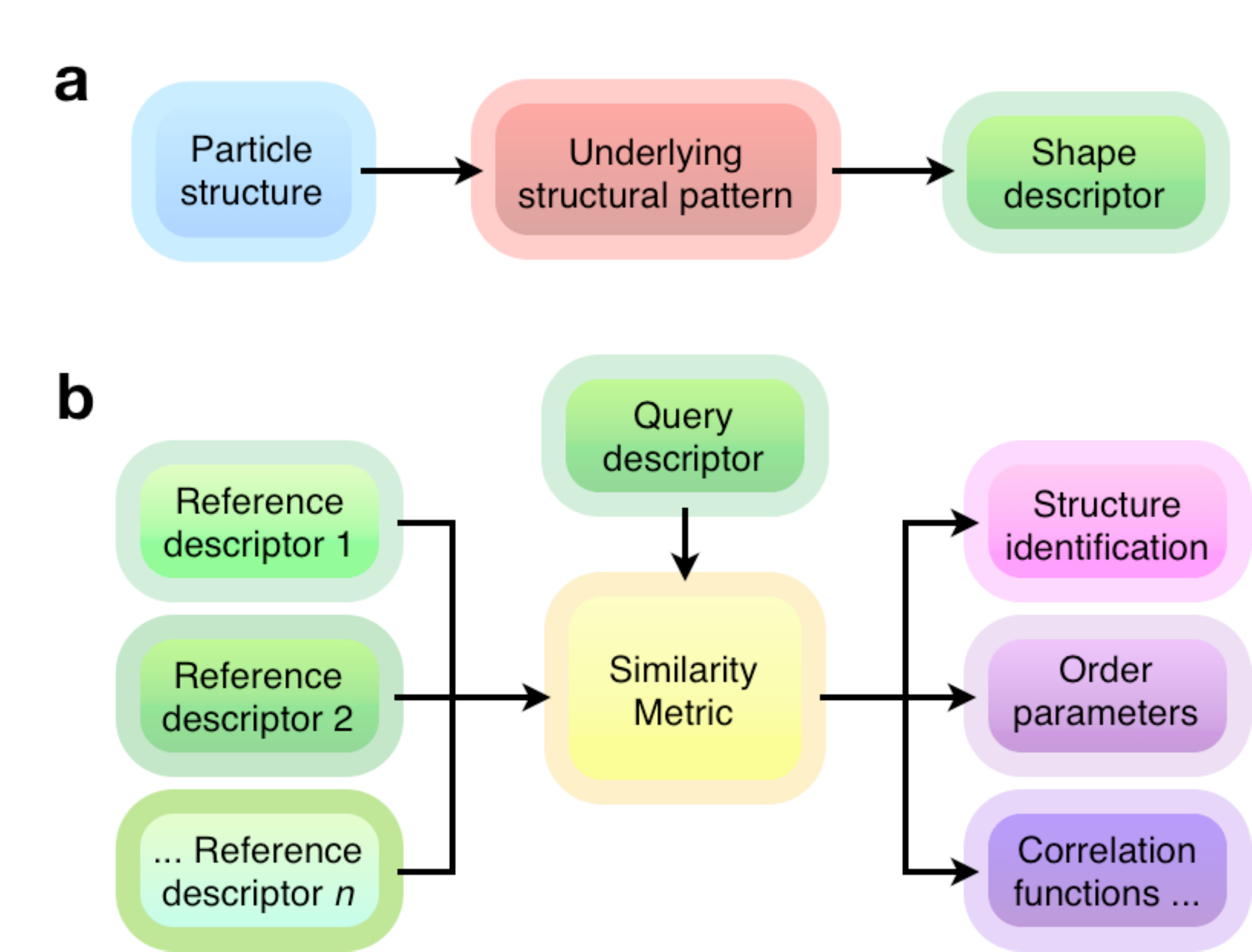}
\caption{Data flow diagram for shape matching. (\textit{a}) A structural pattern is extracted for a given query structure and then indexed into a shape descriptor, which represents a numerical fingerprint for the structure.  (\textit{b}) The shape descriptor is then compared with shape descriptors for reference structures to give a measure of similarity between shapes.  Depending on how we choose the query and reference structures, the similarity value obtained may be applied to constructing order parameters, correlation functions, or other applications.}
\label{fig:dfd}
\end{center}
\end{figure*}

Quantifying how well structures match has been generalized in the context of shape matching\cite{veltcamp} (see Fig.~\ref{fig:dfd}).  Familiar applications include matching fingerprints and signatures\cite{veltcamp},  facial recognition\cite{lu2004} and medical imaging\cite{tagare}. Shape matching defines the concept of the {\it shape descriptor}, a numerical ``fingerprint'' that describes a pattern or shape.  Shape descriptors are associated with {\it query} structures and compared with {\it reference} structures.  The degree of matching between query and reference structures is quantified by a {\it similarity metric}.

Matching information can be used to create order parameters and correlation functions, identify structures, and perform many other types of structural analysis.   Since we can choose virtually any structure as a reference for comparison, shape matching facilitates the creation of highly specific structural metrics.  In the following sections, we review the process of constructing a customized structural metric which involves choosing interesting structures to characterize, computing shape descriptors, and using similarity metrics to compare them.

\subsection{Representative Structural Patterns}
\label{ssec:global}

\begin{figure*}[h!]
\begin{center}
\includegraphics[width=0.82\textwidth]{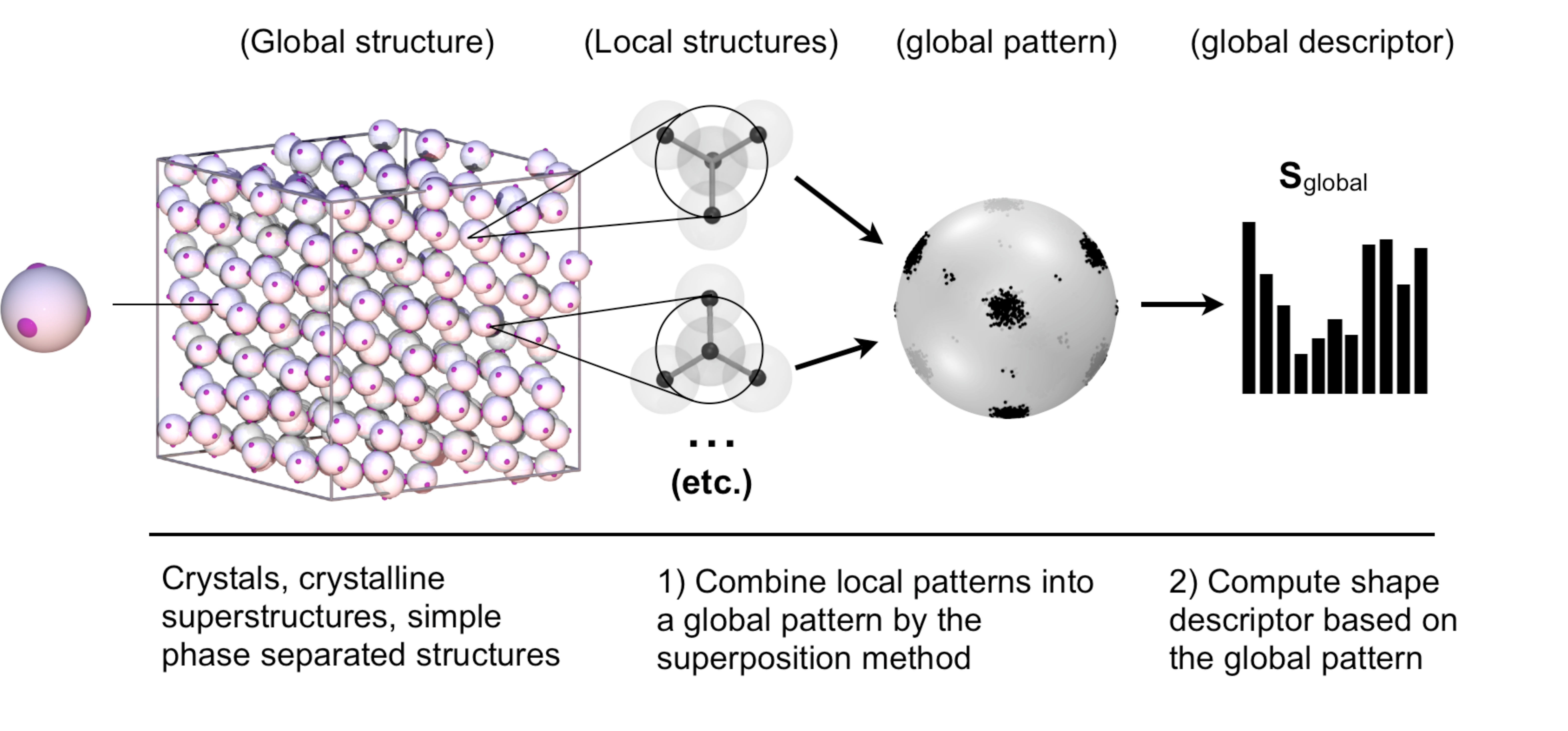}
\caption{Extracting global patterns using the superposition method.   For structures with long-range orientational ordering, such as the diamond structure formed by tetragonally patterned patchy spheres depicted in the panel\cite{zhenlidiamond}, a global pattern is extracted by translating all local clusters\cite{roth2000} or density maps to a common origin.  Here, the local structures are represented by particle positions, but more complex representations are possible.  A global shape descriptor is then computed for the resulting finite structure.}
\label{fig:global1}
\end{center}
\end{figure*}

Before we can compute a shape descriptor, we must extract a representative structural pattern from the system.  This step relies largely on physical intuition; often redundant or unimportant structural information can be discarded out-of-hand to ensure that the matching scheme is only sensitive to important structural features.  One standard type of coarse-graining that is often employed, particularly to the case of small clusters of roughly spherical particles, is to consider particle positions exclusively, discarding information regarding particle sizes and shapes which may be nearly identical (see Fig.~\ref{fig:global1}).  This type of coarse-graining can also be applied to more complex morphologies, such as structures assembled from polyhedral building-blocks\cite{amir09}, or hierarchical assemblies such as micellar systems\cite{iac07, tnv, horsch2005} or virus capsids\cite{virus, rapaport2004self}, wherein the building blocks assemble into larger structural sub-units that arrange into superstructures.  In such cases, the representative structural pattern is given by the positions of assembled sub-units, rather than the individual building blocks (detailed in Example 1 in section~\ref{sec:applications}, below).

\begin{figure*}[h!]
\begin{center}
\includegraphics[width=0.82\textwidth]{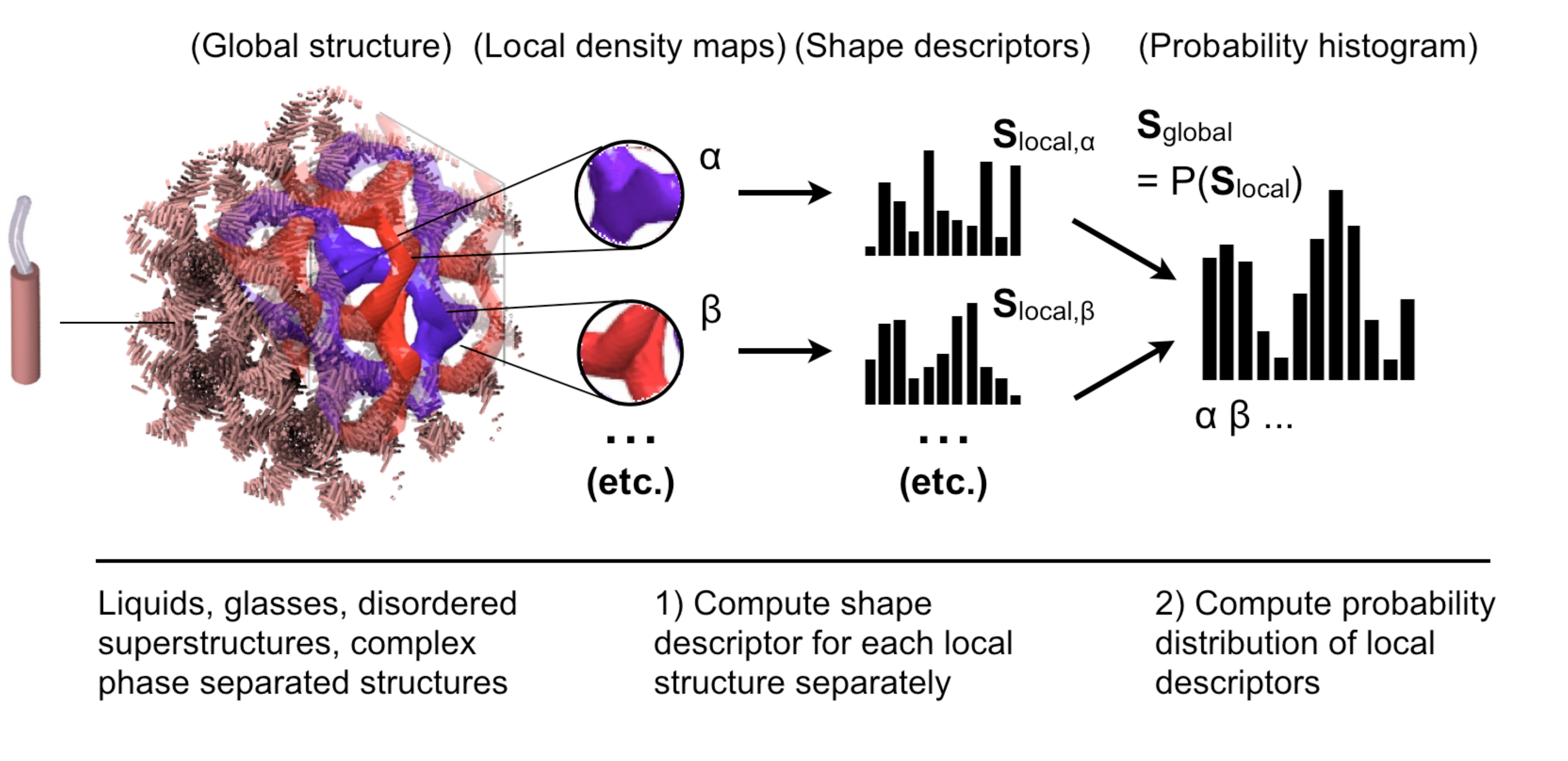}
\caption{Extracting global patterns using the probability distributions method.  For structures without long-range orientational ordering, or complex global structures with many different characteristic directions, a global pattern can be built up from the probability distribution of local patterns.  The double gyroid formed from tethered nanorods\cite{gyroid}, which falls into the latter category, is characterized by computing the distribution of local nanoparticle density maps sampled throughout the structure.  The red/blue color scheme emphasizes the bicontinuous nature of the interpenetrating network.}
\label{fig:global2}
\end{center}
\end{figure*}

Many complex structures cannot be described by positions alone, and require information regarding building block sizes, shapes and orientations.  Such structures can be described by ``volumetric data,'' or ``voxel data'' (i.e., $d$-dimensional pixel data), which is represented numerically by a collection of weights or pixel intensities for cells in a grid that spans space.  This representation is particularly apt for describing the microphase-separated morphologies assembled from systems of tethered nanoparticles and block copolymers, where spatial density maps for the aggregating species may resemble sheet-like or network domains\cite{horsch2005, iacovella2005, tnv, gyroid} (see Fig.~\ref{fig:global2}).  Voxel data captures the essential structural features of these systems, whereas a pattern based on the positions of individual particles within the superstructure does not.  The same rule applies to many other types of structures for which the bulk shape is more important than the underlying particle positions, including all types of phase-separated structures, many complex biological structures such as proteins and macromolecules\cite{mak, grandison}, and large but finite (aka "terminal") nanoparticle assemblies.  Shape descriptors are typically sufficiently flexible to use either voxel data or point cloud data as an input.

Shape descriptors are typically constructed to describe finite objects.  Thus, when describing global structures such as crystals or bulk disordered systems, local shapes must first be extracted from the infinite system and then combined into finite local patterns that reflect the ``global pattern'' for indexing.  The types of global patterns that we create depends on the structural properties of the system.  For structures with long-range orientational ordering, such as crystals and quasicrystals\cite{quasicrystals}, the shape and spatial orientation of local clusters within the system are highly correlated.  Thus, a global structural pattern can be obtained by translating all local shapes to a common origin\cite{snr83}, a scheme that we denote as the ``superposition method.''  The visual depiction of the superimposed structures is known as a ``bond order diagram\cite{roth2000},''  an example of which is depicted for the diamond structure formed by patchy particles\cite{zhenlidiamond} in Fig.~\ref{fig:global1}.   For crystals with multiple particle types, independent global descriptors can be created for each type independently, and a combined descriptor can be created.  Global descriptors based on orientational ordering are applicable to crystalline structures in general, including phase-separated systems arranged in crystalline superstructures\cite{ditethered, iacovella2009b}, where the neighbor directions are computed for the centers of the micelles, cylinders, etc. rather than the individual particles.  Some non-crystalline globally-ordered microphase-separated structures, such as layered or network structures, can be described by superposition as well, where global patterns are built up from local density maps, rather than from local point clusters.  This reflects the fact that the probability density of observing particles in particular spatial directions within these morphologies is often non-uniform.

For systems without long-range orientational ordering such as liquids, glasses and amorphous solids, a different strategy must be employed, since, in such cases, the superposition of local structures inherently yields a uniform pattern.  In such cases, rather than combining neighbor directions or density maps by superposition, we compute a probability distribution of local patterns.  The probability histograms for different structures can then be compared to obtain a measure of similarity between global structures (Fig.~\ref{fig:global2}).  Computing probability distributions is also useful for certain complex orientationally-ordered structures, for which the superposition of local density maps becomes non-distinguishing due to the presence of many different characteristic directions within the structure.  An example of such a structure is given by the double gyroid structure composed of tethered nanorods\cite{gyroid} shown in Fig.~\ref{fig:global2}.

\subsection{Shape Descriptors}
\label{sec:descriptors}

Once we have extracted a representative structural pattern from our particle system, we can compute a shape descriptor to represent the pattern numerically.  Depending on the intended application, different shape descriptors may be best suited to describe a particular structural pattern, and this information should be considered when deciding which shape descriptor to compute.  Below is a short list of desirable shape descriptor properties within the context of assembled systems:

\begin{itemize}
\item  \textbf{Robustness:} the degree of sensitivity to structural defects or random thermal noise.  Some shape descriptors have an inherent data-smoothing mechanism, whereas others require preprocessing to effectively process thermal data.
\item  \textbf{Invariance:} the ability for shape descriptors to remain invariant (i.e., unchanged) under certain mathematical transformations.  Invariance under scaling and translations is typically desirable.  Additionally, descriptors may be invariant under rotations, mirroring operations, or similarity transformations.  Rotation invariance is the most important of these properties for particle systems.  For descriptors without rotation invariance, we often must align or ``register\cite{icp, pca}'' objects prior to matching.
\item  \textbf{Efficiency:} the computational effort required to calculate the descriptor.  For certain applications, CPU time and memory costs may be a limiting factor for choosing a shape descriptor.  For example, efficiency may be an important factor for on-the-fly order parameter calculations that occur during a molecular simulation, whereas for offline data analysis it may be irrelevant.  Often, there is a direct tradeoff between computational cost and accuracy.
\item  \textbf{Comparability:} the ease of matching.  Shape descriptors should yield similar results for similar structures and different results for different structures.  Shape descriptors should be constructed such that similarity is easy to quantify.  The numerical similarity should directly reflect the physical similarity between the shapes used to construct the descriptors. 
\end{itemize}

Below, we review some shape descriptors from the computer science shape matching literature.  Since shape matching is a broad field, we focus on the subset of methods that are best suited for assembled systems.  For a general review of some relevant shape matching methods, see references~\cite{iyerreview, tangelder, zhangreview}.

\begin{figure*}[h!]
\begin{center}
\includegraphics[width=0.75\textwidth]{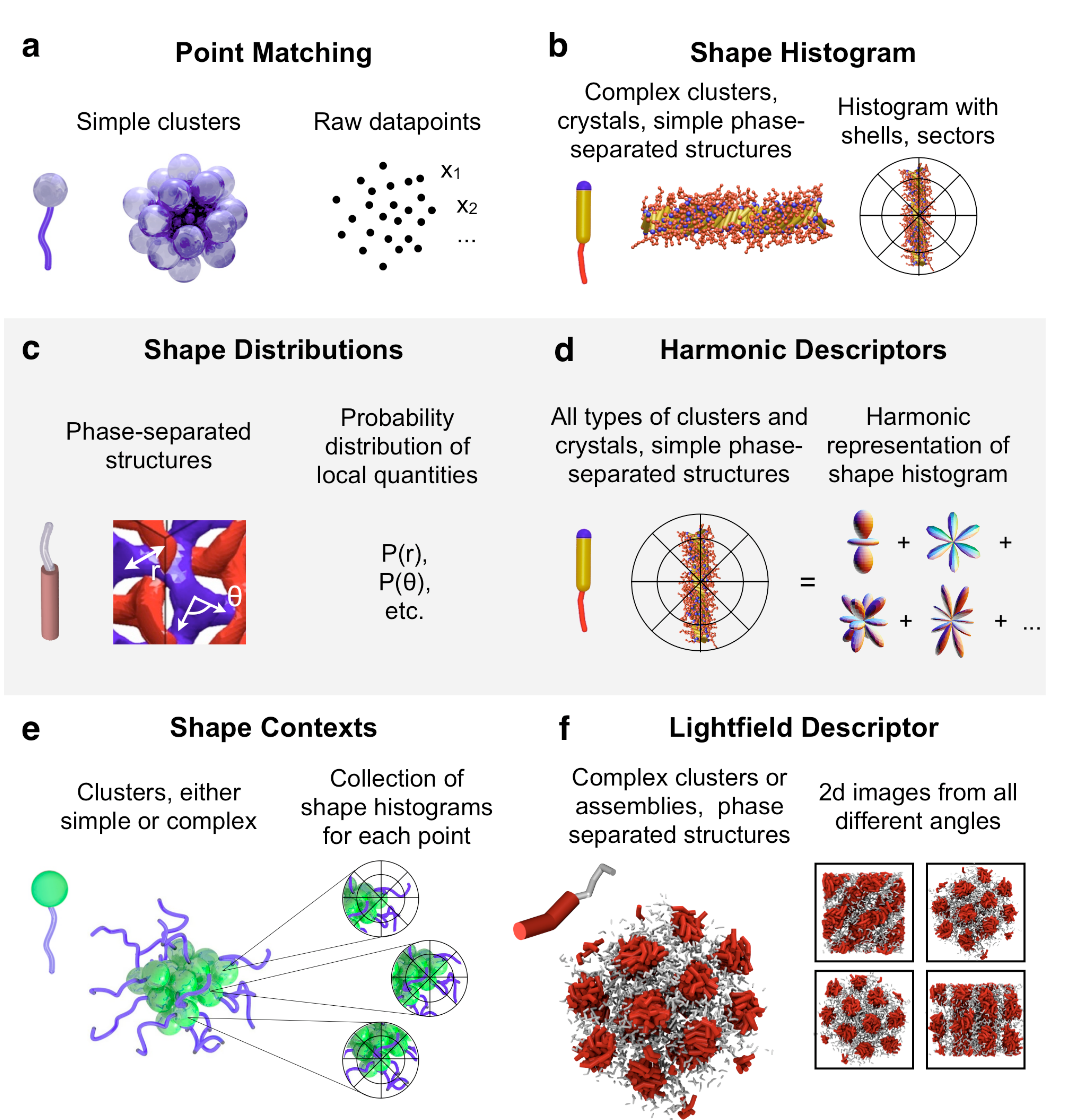}
\caption{Depiction of six different shape descriptors applied to self-assembled systems. (\textit{a}) The point matching descriptor\cite{nrpointmatching, icp}.   Descriptor components are given trivially by particle positions or density maps.  (\textit{b}) The shape histogram descriptor\cite{ankerst}.  The structure is indexed into a spatial histogram consisting of shells and sectors.  (\textit{c}) Shape distribution descriptors\cite{osada}.  The probability distribution is computed for various local measurements, such as the distance or angle between surface points.  (\textit{d}) Harmonic descriptors\cite{fourier, ylm, zernike2d, zernike3d}.  The shape histogram is decomposed into a convenient harmonic representation, which can be used for rotation-invariant matching. (\textit{e}) The shape contexts descriptor\cite{belongie}.  A coarse histogram is created for each point on the structure.  The descriptor is given by the collection of sub-descriptors for each point.  (\textit{f}) The lightfield descriptor\cite{lightfield}.  Images or projections are constructed from several different vantage points and indexed into individual shape descriptors.  The overall descriptor is given by the collection of sub-descriptors for each image.}
\label{fig:descriptors}
\end{center}
\end{figure*}

\bigskip \noindent 
\textbf{Point-Matching Descriptor:} For relatively simple structures such as small clusters of atoms, molecules, or nanoparticle/colloidal building-blocks, we can use the particle positions themselves (or a corresponding density map) as a shape descriptor (Fig. \ref{fig:descriptors}a).  Matching for this scheme is often based on the root-mean-square (RMS) difference between points, and thus the scheme itself is sometimes referred to as ``RMS matching.''  Point matching schemes were applied to early attempts at shape matching for macromolecules\cite{lesk86}, and more complex variations have since been implemented for proteins\cite{proteinpmpnas}.  Point matching schemes have the advantage of being conceptually trivial; however, there are many subtleties associated with these schemes that should be considered.  First, the descriptor requires an assignment step to determine the optimal correspondence between points in compared structures, which is used to re-order the coordinates in the shape descriptors accordingly.  Also, since the descriptors are sensitive to scale, position, and orientation, structures must first be normalized and registered unless the orientations are known beforehand, or rotation-dependent matching is desired.  Depending on the application, shapes may be registered based on rigid alignment, or other constraints.  Since both assignment and registration are computationally expensive (i.e. they scale poorly with the number of points, $n$) point matching descriptors should be avoided unless (1) the number of atoms, molecules, or building blocks that make up the structure is small, (2) matching is required for only a few structures, or (3) registration is not required.

\bigskip \noindent
\textbf{Shape Histogram:} Another conceptually simple shape descriptor that has been applied to molecular database searches is known as the ``shape histogram''\cite{ankerst} (Fig.~\ref{fig:descriptors}b).  This descriptor is based on a density map of the structure on a polar or spherical grid.  Shape histograms are best suited for describing structural patterns that can be broken down into concentric shells, such as nanoparticle clusters, proteins and macromolecules.  Shape histograms are also well suited for indexing global patterns created by the superposition method, as outline above, and can index structures with orientational ordering such as crystals or quasicrystals, and simple microphase separated structures such as layered phases or network structures.  The shape histogram has the advantage over the point matching method that no assignment step is required, since the ordering of points is lost during binning.  Additionally, the grid resolution can be adjusted to provide a variable degree of coarse-graining.   Like the point matching method, the shape histogram requires registration to match non-aligned objects, unless only radial bins are used (i.e., the angular grid resolution is set to zero).   However, shape histograms may lose their discerning capabilities without an angular component.  If $n$ is large, the cost of registration can be significantly reduced by aligning the histograms themselves rather than the underlying structures.  

\bigskip \noindent
\textbf{Shape Distributions:} For many applications, registration is too costly and we require rotation-invariant descriptors.  A simple, yet powerful method for creating invariants, known as the ``shape distributions'' scheme\cite{osada}(Fig.~\ref{fig:descriptors}c), involves computing distribution functions for simple rotationally-invariant local metrics.  Such local metrics are defined based on object surfaces; thus this method is best applied to structures with clearly defined, yet distinguishable, surfaces, such as microphase-separated structures formed by block copolymers\cite{dotera, reister} or tethered nanoparticles\cite{glotzer2005, zhang2003, tns, tnv} (see, for example, Fig.~\ref{fig:global2}).  The shape distribution ``D2'' is defined as the probability distribution of the distance between pairs of surface points.  Another similar distribution ``A3'' is defined by the probability distribution of angles formed by triples of surface points.  Similar distributions are defined for higher numbers of points.  The distributions D2 and A3 are similar to the radial distribution function $g(r)$ and angular distribution function $a(\theta)$, respectively, although usually only surface particles are considered.  Like $g(r)$ and $a(\theta)$, shape distributions are too coarse to distinguish between similar shapes, such as small polyhedral clusters.

\bigskip \noindent
\textbf{Harmonic / Invariant Moment Descriptors:} A more complex, but more powerful method for computing invariant descriptors is to compute the harmonic transform of the shape histogram.  By disregarding the phase information, we obtain descriptors that are invariant under rotations (Fig.~\ref{fig:descriptors}d).  
Like the shape histogram, harmonic descriptors are versatile and can be applied to a wide range of structures including complex nanoparticle clusters, proteins and macromolecules, and crystalline or microphase-separated structures.  The method by which we compute the harmonic transform of the shape histogram depends on the underlying basis.  Invariants can be obtained for shapes on the circle\cite{fourier}($\theta$-dependence), sphere\cite{ylm}($\theta,\phi$-dependence), disk\cite{zernike2d} ($r, \theta$-dependence) and ball\cite{zernike3d}($r,\theta,\phi$-dependence).  On the unit circle or sphere, the harmonic descriptors are called ``Fourier descriptors,'' whereas on the disk or ball, the descriptors are known as ``Zernike descriptors.''  The implementation of these methods for complex assembled systems is described in detail elsewhere\cite{keysSMP-HO}.  
Harmonic descriptors exhibit an inherent data smoothing mechanism that leaves them better-suited for describing small polygonal or polyhedral clusters than the shape histogram, which is prone to error without sufficient averaging.   This property, combined with the property of rotational-invariance, makes harmonic descriptors ideal for describing orientationally-disordered global structures, such as liquids, glasses and certain microphase-separated structures, via the probability distributions method.  Harmonic descriptors also contain additional frequency-dependent information regarding the symmetries of the structure.  These unique properties of harmonic descriptors have already been successfully applied to constructing orientational order parameters for small clusters and simple crystals\cite{halperin, snr83}.

\bigskip \noindent
\textbf{Shape Contexts:}
It is fairly common in the context of self-assembly experiments and simulations to encounter nearly-ideal assembled structures with localized defects.  Thus, it is often desirable to distinguish between local structural dissimilarities that arise due to defects, and ``overall'' differences in the structure.  A brute-force solution to this problem is to explicitly include defective structures in the library of reference structures such that they may be identified directly\cite{iac07}; however, his requires \textit{a priori} knowledge of the entire space of potential defective structures.  Obtaining such knowledge may be intractable for complex assemblies with many degrees of structural freedom, or unmapped systems whose local motifs have not yet been thoroughly studied.  A more general solution is to apply a ``partial matching'' scheme, such as the ``shape contexts'' method\cite{belongie, kortgen2003}, which is capable of matching structures independently of local defects, as well as identifying such defects (Fig.~\ref{fig:descriptors}e).  The shape contexts method combines elements of the point matching scheme with the shape histogram descriptor.  Here, a separate shape histogram is computed for each sample point in the structure, where the coordinate system is centered at that point.  The points in the query structure are then assigned to their corresponding points in the reference structure by optimizing the match between shape histograms.  Outlier points that don't correspond well (i.e., local defects) can be excluded to obtain a partial match, or used to identify the defects.  Shape contexts can be applied to any system where local defects might arise, such as atomic or molecular clusters, micro or nanoscale assemblies, or biological structures.  Since shape contexts are based on the shape histogram, they have the same limitations when indexing structures with a small number of sample points locally.

\bigskip \noindent
\textbf{Lightfield Descriptor:}
The shape contexts descriptor is just one example of the more general method for creating new powerful descriptors by combining simpler sub-descriptors .  A similar method based on combining sub-descriptors is given by the light-field descriptor\cite{lightfield}, which involves projecting 3D structures onto 2D images from 20 vantage points at the vertices of a dodecahedron.  This process effectively simulates the act of viewing a structure from different angles by eye, giving the lightfield descriptor its name (Fig.~\ref{fig:descriptors}f).  The lightfield descriptor can thus be applied to microphase-separated structures, nano/colloidal scale assemblies, or other structures that can be effectively identified by the trained eye.  Each of the 20 2d images is indexed by a 2d descriptor, and assignment is performed for pairs of these descriptors for compared structures to optimize correspondence.  In practice, many initial rotations of the dodecahedron are attempted to minimize error due to small offsets in the spatial orientation. 

\bigskip \noindent
\textbf{Other Possible Descriptors:} 
In addition to the shape descriptors outlined above, the shape matching literature defines numerous potentially useful descriptors that we have not mentioned here.  Some intriguing possibilities include graph based descriptors\cite{topology, skeleton, biasotti}, descriptors based on reflective symmetries\cite{kazhdan2004}, and methods based on the similarity of slices of objects\cite{slice2004}.  Several structural metrics from the condensed matter literature might also serve as useful shape descriptors for some applications.  For example, in the realm of global structures, diffraction patterns, radial distribution functions, or orientation tensors (e.g. the radius of gyration tensor or the nematic order tensor\cite{dijkstra2003}) could be indexed into shape descriptors.  For local structures, analysis schemes such as the common neighbor analysis scheme of reference\cite{cna} could be easily incorporated.  Although many of the structural metrics from the literature may not be independently distinguishing for a wide range of problems, they may still yield useful information as part of a more general scheme through a combination of descriptors.  

\subsection{Similarity Metrics}
\label{sec:matching}

The degree to which two shape descriptors match\cite{veltkamp01} is quantified by a similarity metric.   Computing a similarity metric involves reducing the complex information contained in shape descriptors into a single scalar value that indicates the degree of matching.  The similarity metric that best suits a particular application depends on both the shape descriptor and the intended physical application.  Some desirable properties of similarity metrics are listed below:

\begin{itemize}
\item  \textbf{Metric Behavior:} the ability for a similarity metric to give a value that is proportional to the physical match between the structures.  Some similarity metrics satisfy the triangle inequality\cite{veltkamp01} (i.e., $M(\textbf{S}_A, \textbf{S}_B) + M(\textbf{S}_A, \textbf{S}_C) \geq M(\textbf{S}_B,\textbf{S}_C)$, where $M$ is a similarity metric, and $\textbf{S}_A$, $\textbf{S}_B$, $\textbf{S}_C$ are shape descriptors) and are thus truly metrics, whereas others do not and can be considered pseudo-metrics.  It is typically desirable for similarity metrics to range smoothly with the difference between structures.  
\item  \textbf{Normalization:} the range of possible matching values for a given matching scheme.  For many condensed matter physics applications, we desire similarity metrics that range from 0 to 1 for use as pseudo-order parameters. While many similarity metrics do not vary naturally from 0 to 1, they can often be changed by simply shifting and scaling the interval that defines an ideal and worst-case match.  In practice, there is little difference between this type of pseudo order parameter and a standard order parameter in terms of the underlying physics.
\item  \textbf{Specificity:} the degree to which a similarity metric highlights specific differences between shape descriptors.  For some applications it is desirable to give more weight to specific important differences between the descriptors.  
\end{itemize}

Often, similarity metrics are based on simple geometric functions, such as the Euclidean distance or vector projection between shape descriptors, which are typically represented as long vectors.  Whereas similarity metrics based on the Euclidean distance are particularly common in the shape matching literature\cite{veltcamp}, schemes based on the vector projection are more commonly (implicitly) applied throughout the condensed matter literature\cite{halperin, snr83, tenwolde96}.  In practice, the mathematical form of the similarity metric is typically of little consequence; virtually any function can be chosen, provided it ranges smoothly as the shapes become physically different.  In some specific cases, specialized similarity metrics are designed to be used in conjunction with particular shape descriptors.  The shape histogram scheme described in section~\ref{sec:descriptors} above utilizes a specialized quadratic form distance function for matching\cite{ankerst}, which accounts for mismatches arising from near-misses that occur due to the discrete nature of the histogram bins.  The $P_2$ Legendre polynomial shown in Fig~\ref{fig:nematic} is an implicit example of a specialized similarity metric, specifically designed to match the angles of rod-like particles with the ideal angle given by the global director\cite{liquidcrystals}.

\section{Example Applications}
\label{sec:applications}

In this section, we demonstrate the application of shape matching techniques to a few representative problems from our studies of self-assembly.  For simplicity, we use the same shape descriptor and similarity metric for all of the examples.  Since our goal here is to demonstrate the basic usage of shape matching techniques, our examples should be considered proofs-of-concept rather than optimal solutions to the problems.  Additional examples of applications of other shape descriptors to self-assembly may be found in References\cite{keysSMP-long,keysSMP-HO}.

\subsection{Prototype Shape Matching Scheme}
For our example problems, we use the 3d Fourier shape descriptor\cite{ylm}, which is the harmonic descriptor defined for patterns on the sphere, $[\theta, \phi]$.  We choose this descriptor because it is closely related to the spherical harmonics bond order parameters introduced by Nelson and coworkers\cite{halperin, snr83}, and thus many readers will already be partially familiar with them. The basic idea behind the 3d Fourier descriptor is to decompose a 3d structure into one or more patterns on the 2d surface of a sphere, and represent these patterns mathematically by computing the discrete spherical hamonics transform (DSHT).  This method of representing a pattern as its harmonic transform is analogous to the way that 1d signals along the perimeter of the circle can be described by their discrete Fourier transform (DFT).  

How we extract the patterns on the sphere depends on how data is represented.  For simplicity, we use a minimal data representation based solely on particle positions (i.e., point cloud data) for all of our examples; however other types of data, such as volumetric data, can also be easily treated by Fourier descriptors.  For our examples, we describe particle structures as patterns on the sphere by (1) translating the structure to the origin, (2) grouping all positions within a radial shell $r_s$ and (3) converting each position $\textbf{x}$ into its angular direction relative to the origin $[\theta(\textbf{x}), \phi(\textbf{x})]$.  This is repeated for all $n_s$ radial shells required to describe the full 3d structure, giving $n_s$ patterns on the sphere for each structure.  

For each pattern on the sphere, the Fourier coefficients of the DSHT are given by:
\begin{equation}
\textbf{q}_\ell =  \frac{1}{n} \sum_{i=1}^n  Y_\ell^{m*} [\theta(\textbf{x}_i), \phi(\textbf{x}_i)] \quad m=-\ell, -\ell+1, ...\ell.
\end{equation}
The term $Y_\ell^m$ is a set of spherical harmonics with angular frequency $\ell$.  The coefficients $\textbf{q}_\ell$ are vectors with $2\ell+1$ complex components. Although the Fourier coefficients in their complex number form are rotationally-dependent (i.e., their value depends on the spatial orientation of the underlying pattern), we can convert them to their rotationally-invariant form by  computing the magnitude of each coefficient.  The invariant circular coefficients are given by:
\begin{equation}
|\textbf{q}_\ell| =  \sqrt { \frac{4\pi}{2\ell+1}\sum_{m=-\ell}^\ell |q_\ell^m|^2 }.
\end{equation}
The Fourier invariants are positive real numbers.  Although the coefficient magnitudes themselves can be used directly as order parameters\cite{snr83}, incorporating them into a shape descriptor is often more powerful, since we can compare shapes based on a variety of frequencies and lengthscales.  To create a descriptor from the Fourier coefficients, we simply combine the desired $\textbf{q}_\ell$ or $ |\textbf{q}_\ell| $ into a long vector.  For example, a general rotation-invariant shape descriptor that is applicable to patterns on the sphere over a range of symmetries is given by:
\begin{equation}
\textbf{S}^{F3}_{shell_i}  = <|\textbf{q}_{\ell_{min}}|, |\textbf{q}_{\ell_{min}+1}|, ... |\textbf{q}_{\ell_{max}}|>.
\end{equation}
The range of frequencies can be adjusted to obtain a desired level of resolution.  For our examples below, we use $\ell_{min} = 4$ and $\ell_{max} = 12$.  Since each Fourier descriptor describes the pattern for a given shell, we must combine the Fourier descriptors for each shell to describe the overall shape:
\begin{equation}
\textbf{S}^{F3} = <\textbf{S}^{F3}_{shell_1}, \textbf{S}^{F3}_{shell_2}, \dots \textbf{S}^{F3}_{shell_{n_s}}>.
\end{equation}

For our example applications, we will use a simple similarity metric based on the Euclidean distance $| \textbf{S}_{i} -  \textbf{S}_{j}| $ between harmonic shape descriptors:
\begin{equation}
M(\textbf{S}_i, \textbf{S}_j) = 1 -2 \left( | \textbf{S}_{i} -  \textbf{S}_{j} | /  |\textbf{S}_i| + |\textbf{S}_j|  \right).
\end{equation}
This similarity metric is proportional to the Euclidean distance between shape descriptor vectors, but is normalized such that vectors that match perfectly give a value of $1$, while vectors that are perfectly anticorrelated give a value of $-1$.  Vectors with no directional correlation (i.e., that are orthogonal) give a value of $0$.  This normalization allows us to make a clearer analogy between our matching scheme with a typical order parameter; however, only the relative value of the similarity metric is relevant and the normalization is merely a matter of convenience.

\subsection{Example 1: Micellar Crystal Structures}

\begin{figure}[h!]
\begin{center}
\includegraphics[width=0.75\textwidth]{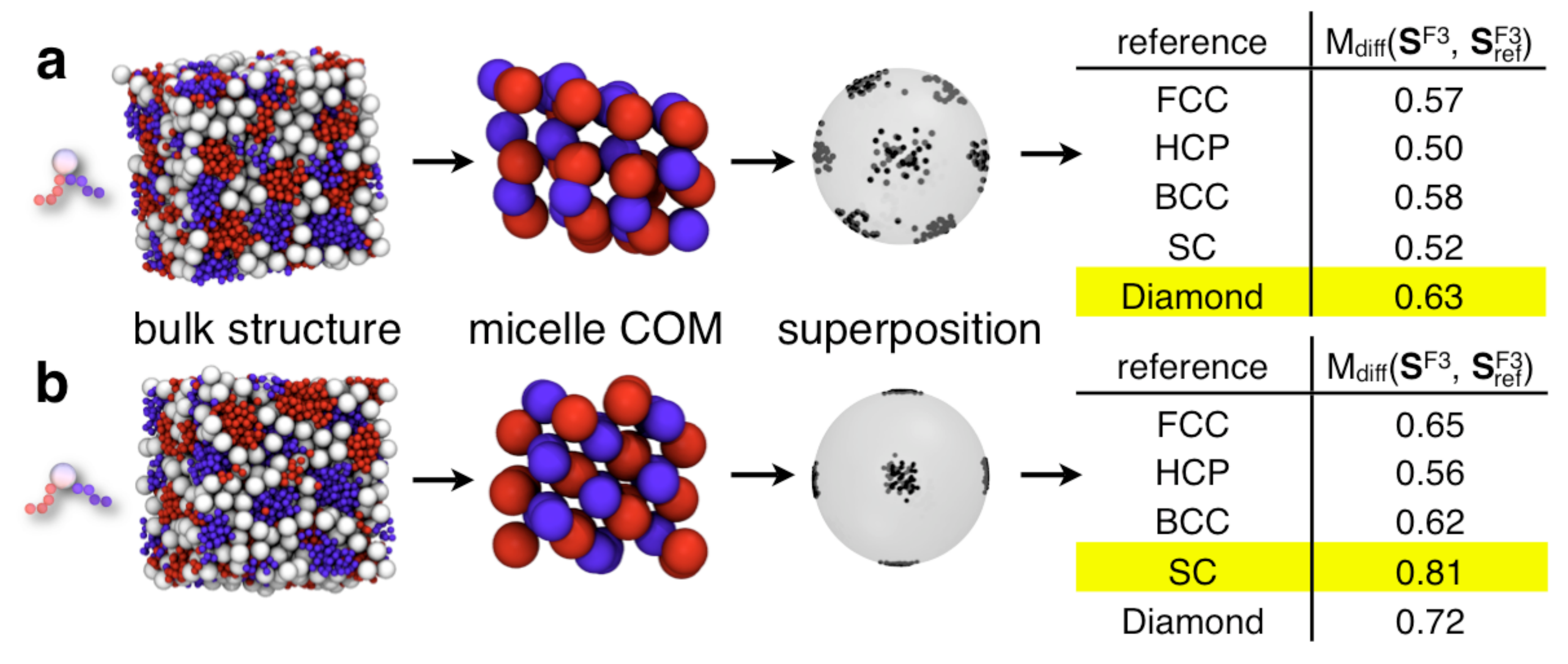}
\caption{Identification of global crystalline structures for a system of ditethered spheres\cite{ditethered, iacovella2009b}. (\textit{a})  A crystal formed in the ditethered nanosphere system where the planar angle between tether attachment is 30 degrees.  Ignoring chemical specificity of the tether micelles, the structure best matches the ideal diamond lattice. (\textit{b}) A crystal formed by the ditethered nanosphere system with planar angle between tether attachment of 60 degrees.  Ignoring chemical specificity of the micelles, the structure best matches an ideal simple cubic structure.  In both cases, the micelle centers are extracted using a Gaussian filter, and matching is based on the global superposition of local patterns (section~\ref{ssec:global}).} 
\label{fig:ditnsmatch}
\end{center}
\end{figure}

A straightforward application of shape matching techniques to particle systems is to identify unknown structures by searching a database of known reference structures.  Structures are identified by the known structure that gives the best match.  Structure identification can be performed for either local structures or for global samples.  As a simple example of structure identification for a global sample, consider the ditethered nanosphere system of references\cite{ditethered, iacovella2009b}, which microphase separates into spherical micelles.  The micelles themselves pack into an ordered binary crystalline superstructure.  Depending on the state point, the system forms different crystals, as shown in Fig.~\ref{fig:ditnsmatch}.  The structural pattern that represents the different crystals is obtained by identifying the micelle centers of mass, which comprise the set of positions that describe the system.  The micelle centers of mass are determined by creating a density map (i.e., a voxel representation) for the aggregating polymer tethers and then applying a Gaussian filtering algorithm adapted from the colloidal science literature\cite{varadan2003, crocker1996} to identify the spheroid centers.  Since the superstructure has long-range orientational ordering, a global pattern is given by the superposition of local patterns (see Fig.~\ref{fig:global1}).  The pattern for the unknown crystal is compared to those for several standard candidate crystals.  For each pattern we compute the 3d Fourier descriptor $\textbf{S}^{F3}$ described above, with rotationally-invariant coefficients for a single shell, $n_s = 1$.  Using this method, the patterns are compared independently of spatial orientation over a single length scale used to construct the local clusters.  The unknown crystal is identified by the reference structure that gives the best match.  The structures in Fig.~\ref{fig:ditnsmatch}a,b are identified as diamond and simple cubic, respectively, where we do not consider the chemical specificity of the two types of micelles. Notice that the best match does not necessarily give a value that approaches $1$; such deviations are common when comparing thermal systems to mathematically perfect reference structures, as we have done here.  The micellar system under investigation exhibits thermal disorder as well as polydispersity in the shape and size of the micelles, and thus particle positions deviate from the ideal lattice points.  Oftentimes, comparing to reference systems that exhibit similar levels of noise may provide clearer results.

This type of database search has already been applied to particle systems in the context of proteins and macromolecules\cite{lesk86, ssm, proteinpmpnas, yeh, venkatraman, mak, grandison}.  Although database searches have only been applied in limited cases to assembled systems\cite{iac07, gyroid}, many standard local structure identification schemes in the condensed matter literature bear a strong resemblance to shape matching identification schemes.  For example, the common neighbor analysis (CNA) scheme of reference~\cite{cna} involves constructing numerical fingerprints for pairs of atoms based on their local neighbor configurations, and identifying local clusters by matching the distribution of fingerprints with those for ideal structures.  In the language of shape matching, the collection of CNA fingerprints can be considered a shape descriptor, and the catalogue of ideal fingerprints can be considered a database of reference structures.  A similar identification scheme is given by the bond order parameters of reference~\cite{snr83}.  Here, particular local structures with strong symmetries, such as small ordered clusters of spherical particles, can be identified by finding structures with bond order parameters that exceed a particular threshold\cite{gasser}.  In this case, the bond order parameters represent shape descriptors, and the threshold values act implicitly as similarity metrics, since the ideal structures are known to have high values of the bond order parameters.

\subsection{Example 2: Icosahedral Clusters of Tetrahedral Particles}

As mentioned in the previous example, a common application of structural characterization schemes is to identify local motifs within a global system.  Examples include finding locally stable clusters in liquids\cite{cna, amir09}, colloids and gels\cite{q6gels} and nanoparticle superstructures\cite{iac07,gyroid}, and identifying structural defects in, or grain boundaries between, crystalline domains, such as in dense colloids\cite{tesfuv2}.  Often, these local structural characteristics can be directly related to the thermodynamic, mechanical, or other properties of the system.

When detecting local structures in systems without long-range orientational order (i.e. ``disordered'' systems), we often encounter structures that are not present in our reference library.  A  structure that does not match with those in the reference library within a certain threshold is considered ``disordered,'' or unimportant\cite{iac07,gyroid}.  The threshold must be chosen carefully; in thermal systems, an overly-stringent cutoff value might cause a matching scheme to miss highly-ordered structures perturbed slightly from their ideal configurations, whereas an overly-permissive cutoff can misidentify highly disordered structures.  In most cases, a sufficiently rigorous cutoff can be defined such that its value does not affect the qualitative results. 

\begin{figure}[h!]
\begin{center}
\includegraphics[width=0.8\textwidth]{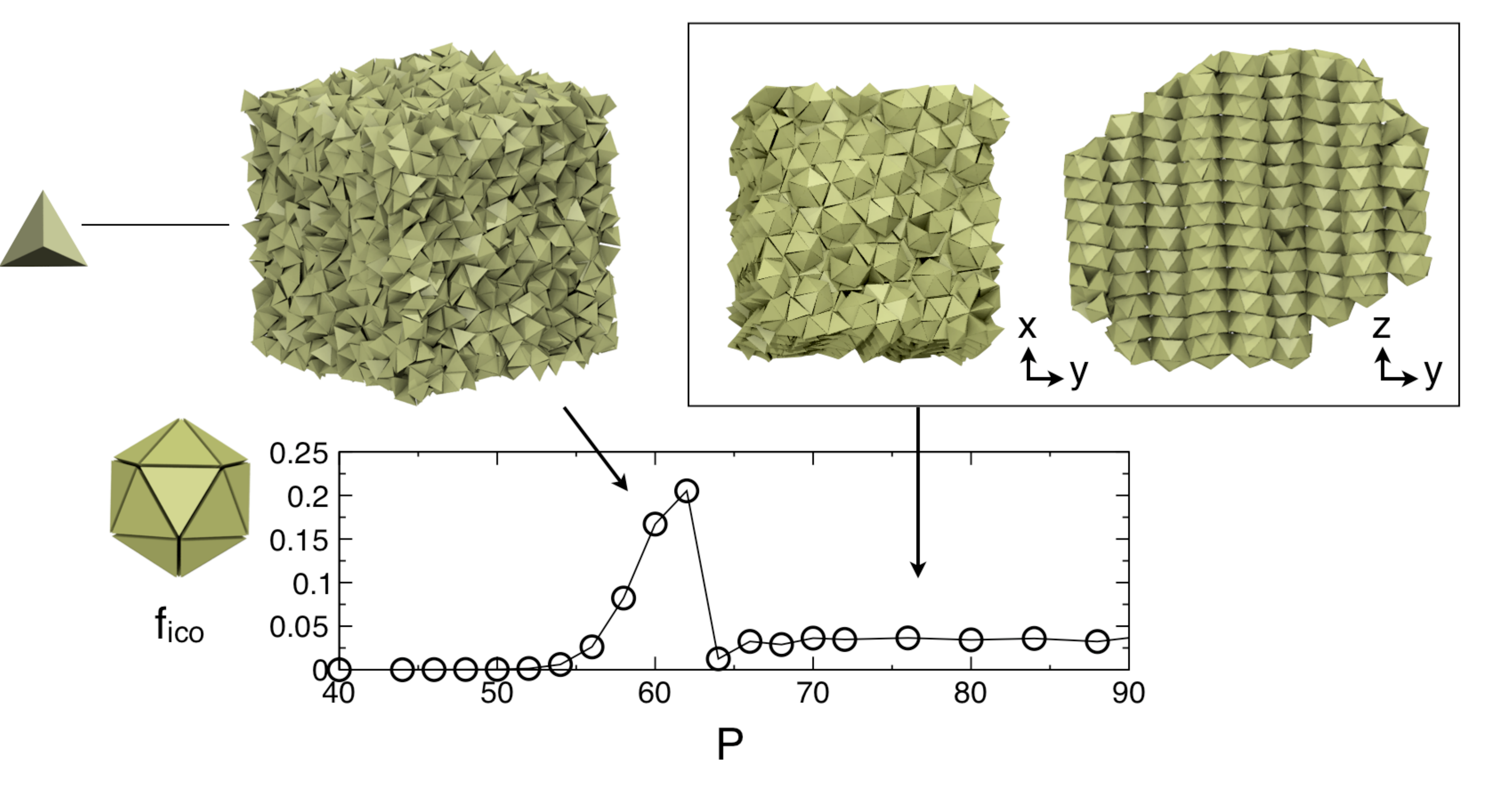}
\caption{Icosahedral clusters in the hard tetrahedron system\cite{amir09}.  As the pressure and the corresponding density increase, icosahedra grow more prevalent until the system transforms into a dodecagonal quasicrystal at $P \approx 62$, at which point the number of icosahedra vanishes. }
\label{fig:tetrahedra}
\end{center}
\end{figure}

As an example of identifying ordered local structures in an otherwise disordered system, consider the hard tetrahedron fluid studied in reference~\cite{amir09} (Fig.~\ref{fig:tetrahedra}a).  In this system, an important local motif to both the fluid and the glass, originally identified by visual inspection, is the icosahedron formed by 20 tetrahedra sharing a common vertex.  To identify icosahedra in the system, we first cluster all sets of 20 tetrahedra in the system that share a common vertex.  The structural pattern for each cluster is defined by the directions of vectors drawn from the center of the cluster through the face of each of the 20 tetrahedra, which for an ideal icosahedral cluster results in a dodecahedron.  Any local cluster $i$ that matches the shape of a dodecahedron with a value of $M_{cut}(\textbf{S}^{F3}_i, \textbf{S}^{F3}_{dodecahedron}) > 0.9$ is considered to be in an icosahedral motif.  Fig.~\ref{fig:tetrahedra} shows the fraction of tetrahedra that participate in at least one icosahedron as a function of pressure. Icosahedra are relatively common in the tetrahedral fluid (below $P=62$) and become more prevalent with increasing density, persisting into the glass if the fluid is compressed too quickly.  As the fluid transforms into a quasicrystal at $P\approx 62$, the fraction of tetrahedra in icosahedra decreases drastically, and vanishes for the ideal quasicrystal without thermal fluctuations.   Although the value of $M_{cut}$ may affect the absolute number of icosahedra, the same underlying physical transition is captured for any reasonable value.

\subsection{Example 3: Assembly of a Helical Ribbon}

Another standard application of structural metrics is to track structural transitions, either as a function of time or a changing reaction coordinate.  This is typically accomplished by monitoring either an order parameter or correlation function as the system goes through a transition.  Tracking structural transitions is important for a wide variety of applications, including elucidating thermodynamic transitions\cite{tenwolde96,  halperin74, smectic, gubbins, confinedfluids} and assembly pathways\cite{zhenlidiamond, yamaki, bladon93, tang2006}.  Many of the advanced molecular simulation techniques used to study transitions\cite{umbrella, tps,tpsropes, ffs, metadynamics} rely on structural metrics in the context of pseudo-reaction coordinates\cite{tps}, biasing parameters\cite{umbrella}, and collective variables\cite{metadynamics} to guide the statistical sampling algorithm.  Standard order parameters have been devised for various types of ordering, including bond orientational ordering\cite{halperin78, nelsonc6, snr83, steinhardt81}, liquid crystalline ordering\cite{liquidcrystals, larson} such as nematic\cite{nematic} and smectic\cite{smectic} phases, chiral ordering\cite{kamienchiral}, and helical ordering\cite{h4}.  Time correlation functions based on these types of order parameters have been applied to creating structural ``memory'' functions for glassy liquids\cite{kawasaki, tanaka} and ordered motifs attaching to a growing quasicrystal nucleus\cite{keys07}. 

\begin{figure}[h!]
\begin{center}
\includegraphics[width=0.8\textwidth]{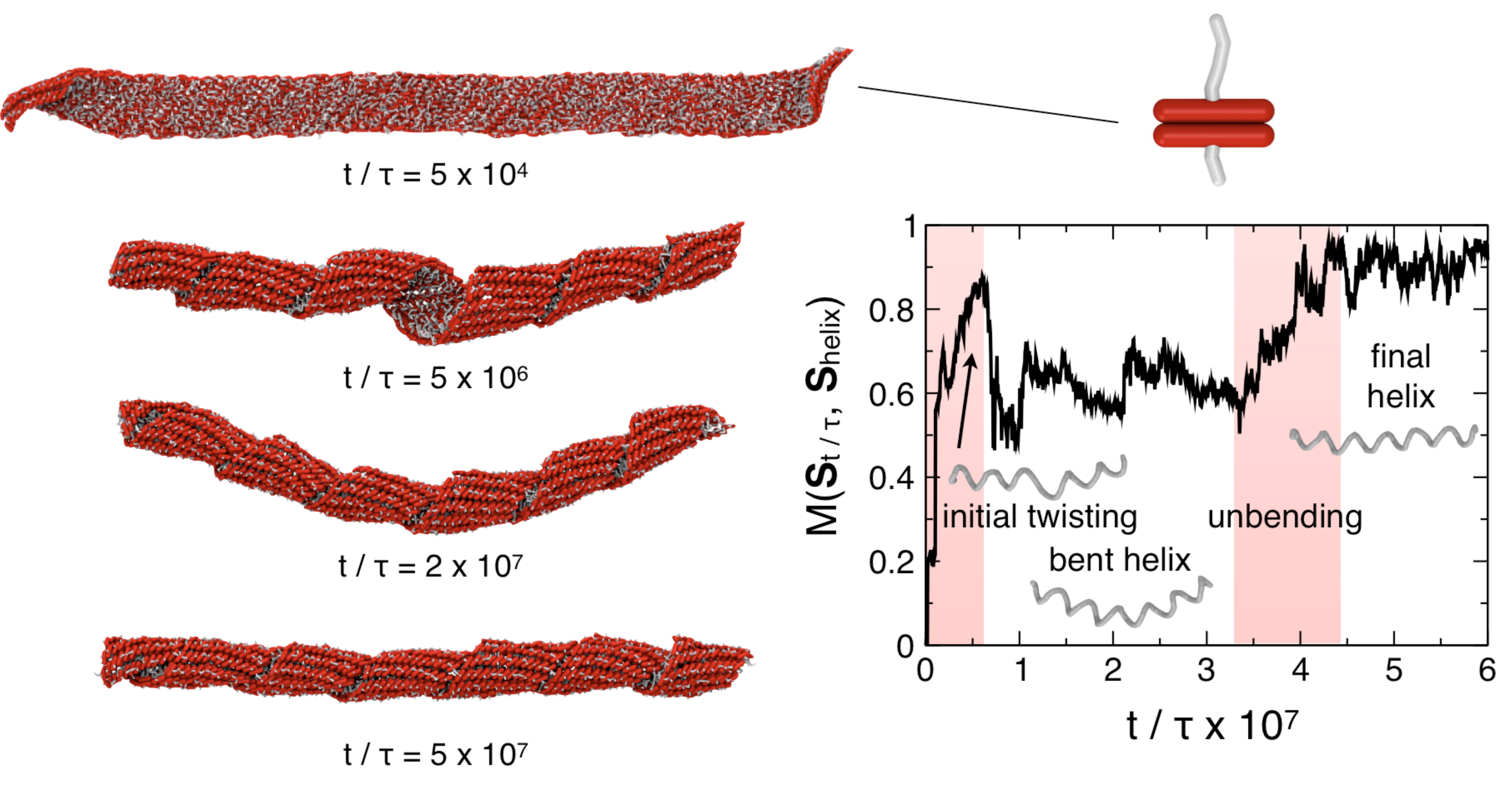}
\caption{Assembly of a helical sheet composed of laterally tethered nanorods\cite{trunghelix}.  The rods form a bilayer with long attractive tethers on one side, and shorter attractive tethers on the other.  As time progresses, the sheet folds into a helix to maximize the favorable energetic interactions between the longer tethers.  The matching order parameter $M_{dist}(\textbf{S}_{t / \tau}, \textbf{S}_{helix})$ compares the structure at time $t$ with the shape of the final ideal helical structure.}
\label{fig:helix}
\end{center}
\end{figure}

As a simple example of using shape descriptors to create an order parameter, consider the ribbon-like bilayer composed of laterally tethered nanorods studied in reference\cite{trunghelix}, and shown in Fig.~\ref{fig:helix}.  The initial sheet or ribbon is unstable and eventually relaxes into a stable helical structure.  We can track this structural transition by matching the shape of the sheet at a given time $t$ with the final, fully-equilibrated helical structure: $M(\textbf{S}^{F3}_t, \textbf{S}^{F3}_{helix})$.  Since the structure is 3-dimensional and has radial dependence, we use a Fourier descriptor with $n_s = 6$ radial shells: $r_s = 10\sigma, 30\sigma ... 110\sigma$, where $\sigma$ is the distance unit corresponding to a Lennard-Jones particle diameter.  Since the sheet only changes in terms of its twist in space, we save computational effort by only considering points along the backbone of the sheet.  Fig. \ref{fig:helix}a shows the helical order parameter as a function of time for a long molecular dynamics run.  We observe that the sheet begins to twist from both ends simultaneously, which gives rise to a defect at the center of the helix, where a mismatch in the periodicity between the two ends occurs.  This results in a tendency for the structure to bend to close the defect.  The bend persists for many millions of time steps before annealing into a defect free helix at around $t/\tau = 4.5 \times 10^7$.  This behavior is well captured by matching the overall shape of the structure, but is not captured by the more standard $H_4$ descriptor, applied in the original reference, which only measures the degree of helical ordering and gives an essentially constant value for all times after the completion of twisting at $t/\tau \approx 7 \times 10^6$\cite{trunghelix}.  Using $H_4$ alone, it would appear that the structure is fully formed at this early time, which does not capture the important defect removal behavior, which can also be observed by visual inspection.  

\section{Future Outlook}
\label{sec:future}

Beyond identifying local and global structures and tracking structural transitions, there are many more applications of shape matching.  In this section, we briefly review some areas in which we are currently applying shape matching for studying self assembly.  Additional details may be found in Ref.~\cite{keysSMP-long} and in the individual references cited below.

\begin{figure}[h!]
\begin{center}
\includegraphics[width=0.75\textwidth]{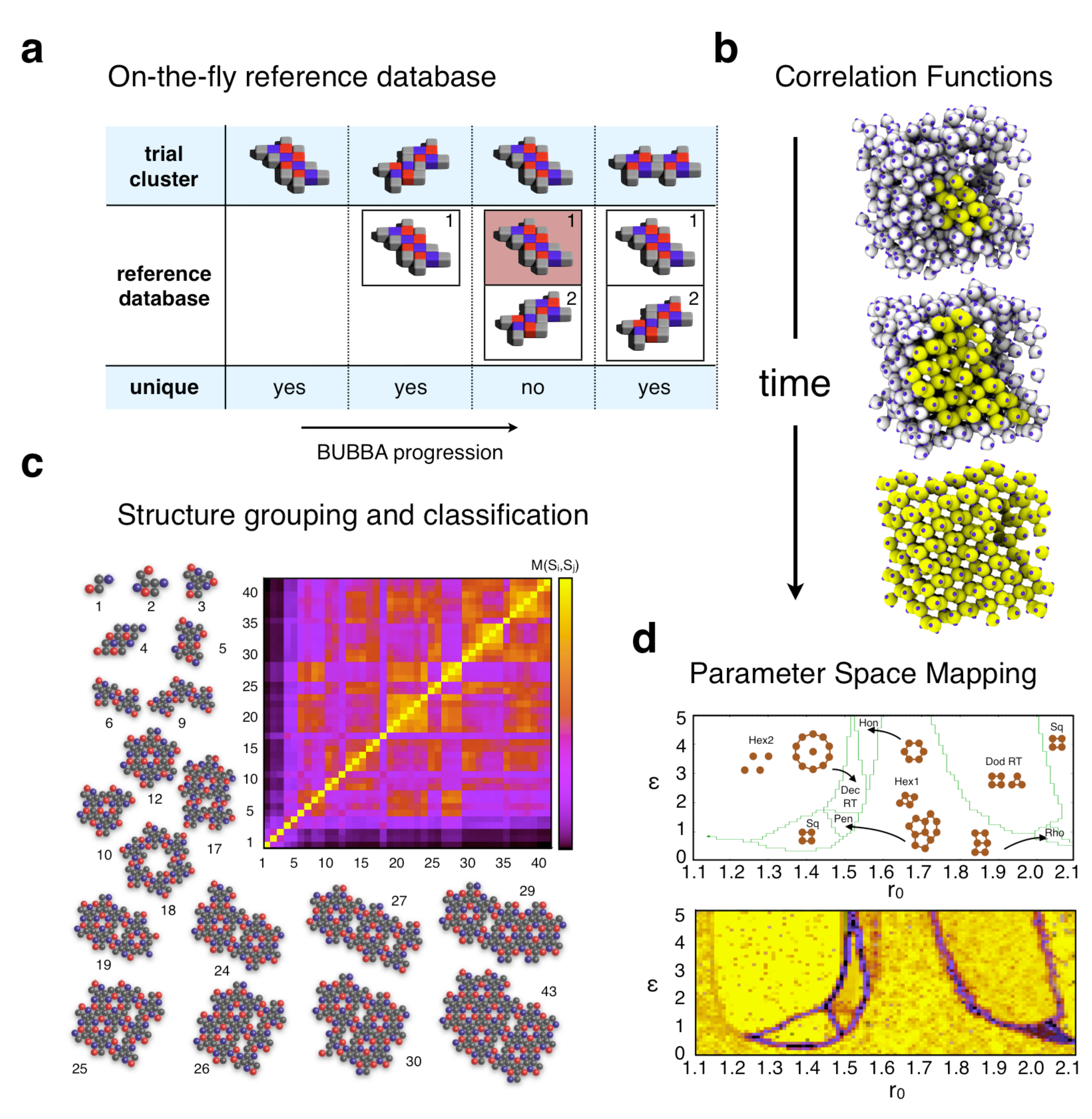}
\caption{Potential uses for shape matching in assembly applications.  (\textit{a}) Searching parameter spaces for unique structures.  The panel depicts the Bottom-Up-Building-Block-Assembly (BUBBA) algorithm\cite{bubba}. (\textit{b}) Computing spatial correlation functions.  The panel depicts detecting a growing diamond crystal nucleus in a system of patchy particles\cite{zhenlidiamond}.  (\textit{c}) Structure grouping and classification.  The panel depicts a similarity matrix (i.e., all of the pairwise similarity values) for 2d clusters of different sizes.  Groups of similar structures are identified by bright boxes about the line y=x.   (\textit{d}) Abstract correlation functions.  The panel depicts a structural phase diagram for the 2d Lennard-Jones Gauss system (top), created by visual inspection\cite{engel07}, compared with a phase diagram for the same system generated automatically using a shape matching algorithm (bottom).}
\label{fig:scope}
\end{center}
\end{figure}

\bigskip \noindent
\textbf{On-The-Fly Structure Identification:} For many assembly applications, such as Bottom-Up-Building-Block-Assembly (BUBBA)\cite{bubba}, we are interested in cataloguing unique structures.  When enumerating unique structures it is not typically necessary (or feasible) to define a library of reference structures \textit{a priori}, as we did for examples 1 and 2 above.  Rather, the reference library can be compiled on-the-fly as new structures are encountered (see Fig.~\ref{fig:scope}a).  Each new structure is given a unique identifier, and structures that are duplicates are labeled with the same identifier.  In addition to cluster enumeration schemes, this type of algorithm can potentially be applied to automatically detect regions of unique ordering in structural phase diagrams.

\bigskip \noindent
\textbf{Space/Time Correlation Functions:} In example 3 above, we demonstrated how shape matching could be used to track a structural transition as a function of time, or a reaction coordinate.  Another common application of structural metrics is to characterize how structures change in space.   In the context of shape matching, this involves choosing structures from different points in the system, rather than ideal structures, as reference structures. Spatial correlation functions are often used to measure structural ``correlation lengths.''  In the condensed matter literature, structural correlation functions have been defined for crystal-like ordering in 2d\cite{halperin78, nelsonc6} and 3d\cite{snr83, tenwolde96}, nematic ordering\cite{allen97}, and many other more specialized types of ordering.  More specialized types of spatial correlation functions have been widely applied as well.  One example is the $q_6 \cdot q_6$ scheme of references~\cite{tenwolde96, auer04}, which detects ordered crystal nuclei based on spatial correlations between local bond order parameters.  This scheme can be adapted to identify crystal nuclei in general by replacing $q_6$, which is only sensitive to particular crystal structures, with other shape descriptors that are applicable to a particular crystal under investigation.  Fig.~\ref{fig:scope}b depicts the formation of a diamond-structured crystal nucleus (yellow) in a system of patchy particles, identified by replacing $q_6$ with the $\ell=3$ Fourier coefficient, $q_3$\cite{zhenlidiamond}.

\bigskip \noindent
\textbf{Structure Grouping and Classification:}  The field of self-assembly involves a wealth of particle building blocks and the assemblies they form; thus  it is sometimes useful to categorize or classify structures based on particular structural features.  For example, reference~\cite{glotzer07} ranks different building blocks for self-assembly based on their shape anisotropy.  Shape matching methods can provide numerical metrics by which to classify structures.  Structures can be ranked based on the degree to which they exhibit a particular structural feature of interest, or by how well they match ideal structures exhibiting a particular feature.  For example, structures can be ranked based on their 6-fold symmetry by computing the value of their $\ell=6$ Fourier descriptor, which is proportional to the degree of 6-fold symmetry.  Similarly, we can create groups of structures that exhibit a particular structural feature by comparing shape descriptors.  One example of a technique used to visually group similar structures is given by plotting a matrix of pairwise similarity values known as a  ``similarity matrix'' or ``heat map''\cite{princetonshape}, as depicted in Fig.~\ref{fig:scope}c for 2d colloidal clusters\cite{ortiz10}.  Groups of clusters with similar structural features produce bright blocks, indicating that clusters within this region of parameter space match well.  Grouping objects based on shape similarity has also been applied recently to macromolecules and proteins\cite{mak, yeh}. 

\bigskip \noindent
\textbf{Abstract Correlation Functions:}  Thus far, we have either extended the applicability of standard condensed matter order parameters and correlation functions by incorporating shape matching, or applied standard shape matching applications directly in the context of assembled systems.  However, in addition to extending existing applications for use with assembled systems, shape matching allows us to invent new methods that have not yet been explored.  For example, rather than creating correlation functions in space and time as we typically do for condensed matter systems, we can create abstract correlation functions in parameter space. Fig.~\ref{fig:scope}d depicts a parameter space correlation function computed for the 2d Lennard-Jones Gauss system\cite{engel07}, which identifies structural phase boundaries (purple) by finding points in parameter space that do not match well with their neighboring points.  This correlation function is able to reproduce the structural phase diagram produced in reference~\cite{engel07} by visual inspection of over 5000 independent configurations.  This scheme is just one example of how shape matching algorithms can replace the human element in searching for target structures, and rapidly mapping parameter spaces.  The ability to expedite self-assembly research by automating the study of unique systems may represent one of the most important uses for shape matching moving forward.   

\bigskip \noindent
\textbf{Summary:} The example applications and shape descriptors that we have provided here represent only a small subset of the vast range of possibilities yet to be explored.  In the future, the wealth of shape descriptors from the shape matching literature should be tested for different classes of particle systems to expand the scope of order parameters available to the fields of experimental and computational assembly.  New abstract order parameters and correlation functions, such as the phase space correlation function of Fig.~\ref{fig:scope}d, can be constructed to expand the algorithms used to explore new systems.  More immediately, the relatively simple algorithms outlined here can be applied to existing assembled systems to enhance our ability to gain insight into the underlying physics of these complex systems.

\textbf{\textit{Acknowledgements:}}  The Department of Education (GAANN Grant No. P200A070538) provided partial support for ASK. The National Science Foundation supported ASK, CRI and SCG in the development of shape-matching reference databases (Grant No. DUE-0532831) and shape-matching computational codes (Grant No. CHE-0626305). The Biomolecular Materials and Processes Program of the Department of Energy (Grant No. DE-FG02-02ER46000) supported CRI and SCG in the application of shape-matching to the system of ditethered nanospheres presented here. CRI also acknowledges a University of Michigan Rackham Predoctoral Fellowship.  We thank T.D. Nguyen, M. Engel, and G. van Anders for helpful comments on the manuscript.  Thanks also to T.D. Nguyen, M. Engel, A. Haji-Akbari, E. Jankowski, C. Phillips, D. Ortiz, A. Santos, C. Singh, A. Mohraz, and M. Solomon for providing example data, not all of which could be used here.

\bibliography{arcmp_shape}
\bibliographystyle{ieeetr} 

\end{document}